\documentclass[longnamesfirst,apjl]{emulateapj}
\usepackage{epsfig}
\usepackage{apjfonts,natbib,epsfig}
\usepackage{apjfonts,natbib}
\usepackage{amsmath}
\usepackage{graphics}
\tighten

\usepackage{color}

\newcommand{\beq}{\begin{equation}}
\newcommand{\eeq}{\end{equation}}
\newcommand{\beqa}{\begin{eqnarray}}
\newcommand{\eeqa}{\end{eqnarray}}

\begin{document}

\righthead{TWO SN POPULATIONS \& COSMOLOGY}
\lefthead{SARKAR ET AL.}

\title{Implications of Two Type Ia Supernova Populations for
Cosmological Measurements} \author{Devdeep Sarkar\altaffilmark{1},
Alexandre Amblard\altaffilmark{1}, Asantha Cooray\altaffilmark{1}, and
Daniel E.  Holz\altaffilmark{2,3}} \altaffiltext{1}{Department of
Physics and Astronomy, University of California, Irvine, CA 92697}
\altaffiltext{2}{Theoretical Division, Los Alamos National Laboratory,
Los Alamos, NM 87545} \altaffiltext{3}{Kavli Institute for
Cosmological Physics and Department of Astronomy and Astrophysics,
University of Chicago, Chicago, IL 60637}

\begin{abstract}
Recent work suggests that Type Ia supernovae (SNe) are composed of two
distinct populations: prompt and delayed.  By explicitly incorporating
properties of host galaxies, it may be possible to target and eliminate
systematic differences between these two putative populations. However, any
resulting {\em post}-calibration shift in luminosity between the components will
cause a redshift-dependent systematic shift in the Hubble diagram.  Utilizing an
existing sample of 192 SNe Ia, we find that the average luminosity difference
between prompt and delayed SNe is constrained to be $(4.5
\pm 8.9)\%$.   If the absolute difference between the two populations is 0.025
mag, and this is ignored when fitting for cosmological parameters, then the dark energy equation of state (EOS) determined from a sample of
2300 SNe Ia is biased at $\sim1\sigma$.
By incorporating the possibility of a two-population systematic, this bias can
be eliminated. However, assuming no prior on the strength of the
two-population effect, the uncertainty in the best-fit EOS is
increased by a factor of 2.5, when compared to the equivalent sample
with no underlying two-population systematic.  To avoid introducing a
bias in the EOS parameters, or significantly degrading the measurement accuracy, 
it is necessary to control the post-calibration luminosity difference between
prompt and delayed SN populations to better than 0.025 mag.
\end{abstract}


\section{Introduction}
\label{sec:introduction}

The discovery of the accelerating expansion of the
universe~\citep{Riess:98, Perl:99} has led to an
explosion of interest in the underlying physics responsible for this
acceleration.  A favored model characterizes the acceleration by an
unknown energy density component, dubbed the dark energy. While there
exist a variety of probes to explore the nature of this dark energy,
one of the most compelling entails the use of type Ia supernovae
(SNe henceforth) to map the expansion history of the
universe. Several present and future SN surveys are aimed at
constraining the dark energy equation-of-state (EOS) to
better than 10\%. With increasing sample sizes, SN distances can
potentially provide multiple independent estimates of the EOS of dark
energy when binned in redshift \citep{Huterer:05,sullivan:07,sarkar:08a}.  Given the importance of
dark energy measurements, it is then useful to quantify various
systematics that impact SN cosmology \citep{huigreene:06,cooraycaldwell:06,cooray:06,sarkar:08b}.

Recently, suggestions have been made that the SN population consists
of two components, with a ``prompt'' component proportional to the
instantaneous host galaxy star formation rate, and a ``delayed'' (or
``extended'') component that is delayed by several Gyrs
\citep{hamuy:95, livio:00, scannapieco:05,mannucci:06, sullivan:06,
strovink:07}.  The former is expected to be more luminous,
and thus, prompt SN lightcurves are broader than those of the delayed
population. By classifying SNe by host galaxy type,
\citet{howell:07} found a {\em
pre}-calibration intrinsic luminosity difference of $\sim (12 \pm 4)$\% between the two components, 
based on a difference of $(8.1 \pm 2.7)\%$ in the width of
lightcurves. Since the SN lightcurves are used to calibrate the
intrinsic luminosity \citep{phillips:93, riess:96, perlmutter:97,
tonry:03, prieto:06, guy:07, jha:07}, a systematic difference in
intrinsic luminosity could conceivably be calibrated out, if
the SN lightcurve calibration relation is the same for both populations.

However, it is unclear whether the full intrinsic difference in
luminosity between the two populations is captured by a calibratable
difference in the lightcurves.  A residual in the calibrated
luminosity could potentially remain, leading to a redshift-dependent
shift in the Hubble diagram, and systematic errors in the best-fit
cosmological parameters. For example, it is likely that the intrinsic colors of
Type Ia SNe are not uniquely determined by light-curve
shape~\citep{conley08}. Even if this is not the case, differences in intrinsic color between
the populations might introduce systematic differences in post-calibration luminosity (e.g., through differences in the
extinction corrections).
We model the two-population systematic, constraining the magnitude of the effect
with current data. With large SN samples it may be possible to estimate the
magnitude of the systematic directly from the data (e.g., by
correlating observed SN brightnesses with properties of the host
galaxies \citep{hamuy:96b,Riess:99,sullivan:06,jha:07,
gallagher:08}). We quantify the level of calibration required to
avoid significantly degrading the determination of the dark energy EOS.

The paper is organized as follows: in $\S$2 we discuss a model for
incorporating a two-population systematic residual luminosity into the
Hubble diagram. In $\S$3 we investigate the possibility of
detecting this systematic from both current and future SN data, and
the impact on dark energy parameter estimation.

\begin{figure*}[!t]
\centerline{
\includegraphics[scale=0.5]{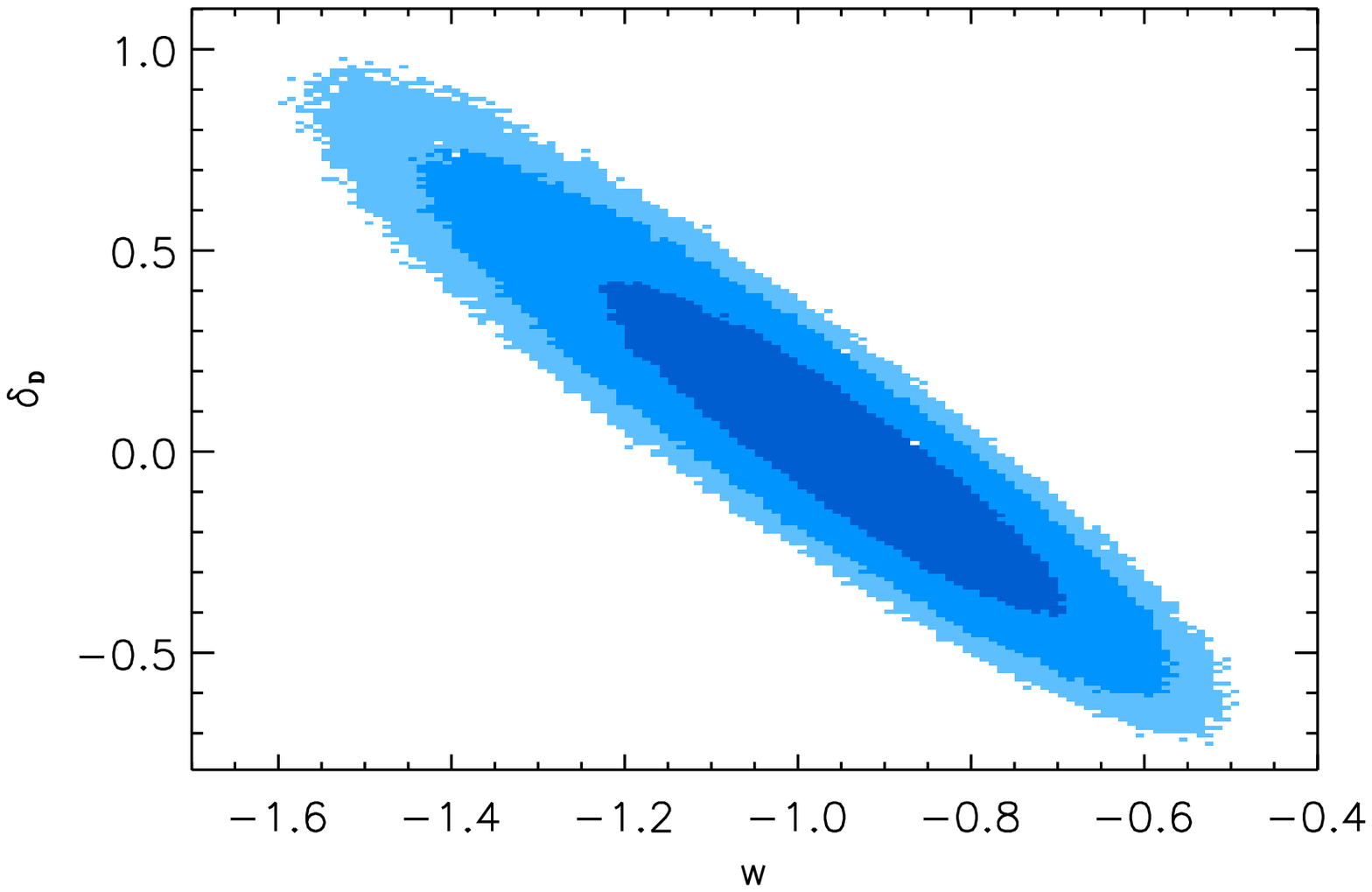}
\includegraphics[scale=0.5]{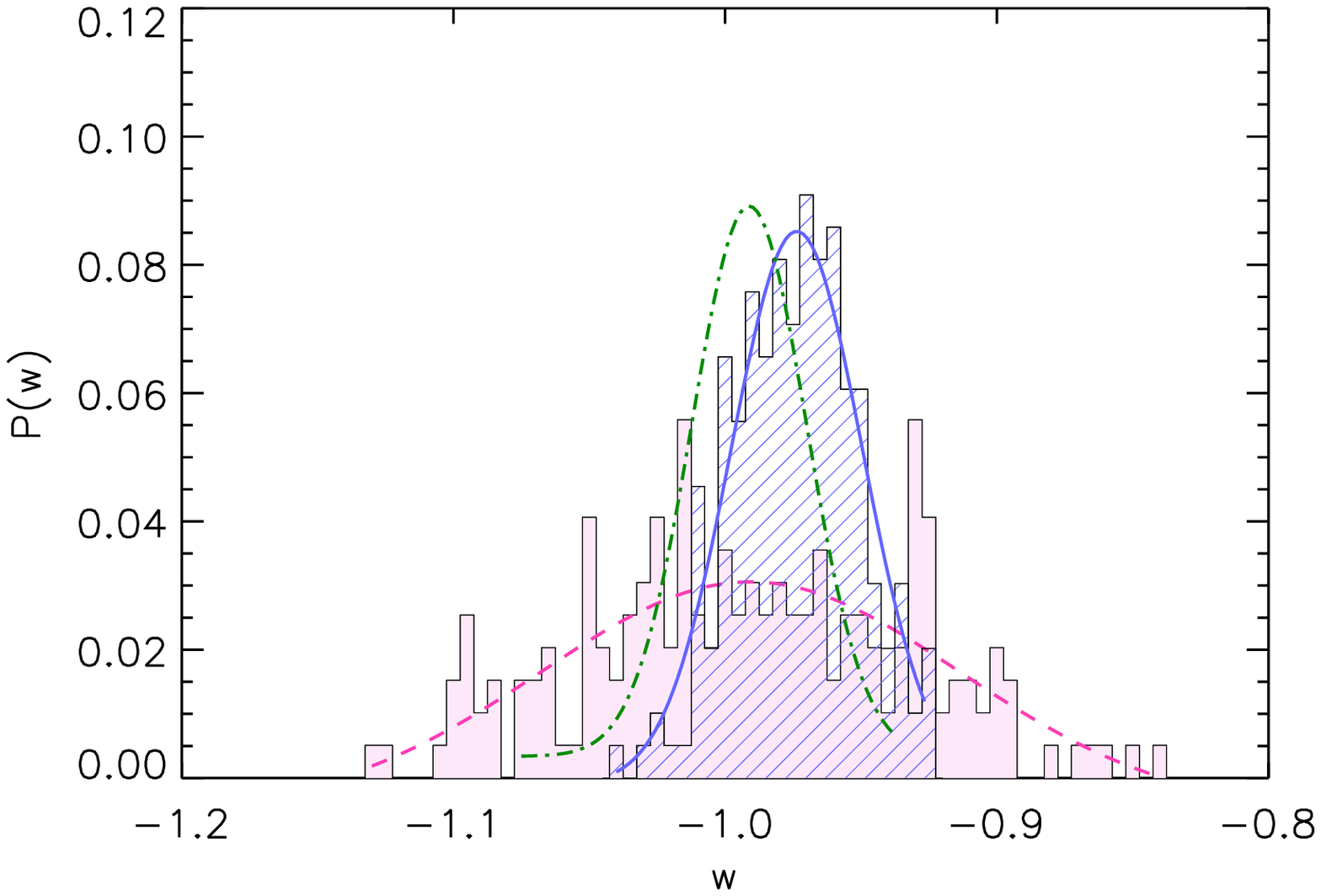}
}
\caption{{\it Left panel:} Correlation between   
$\delta_D$ and $w$, corresponding to a $w$CDM model, for the 192 SN
dataset discussed in the text. There is a strong
degeneracy between the two-population systematic, $\delta_D$, and the dark
energy EOS, $w$. This leads to a potential bias
in the measurement of the EOS, and increased errors when the fits are marginalized over $\delta_D$.
{\it Right panel:} Histograms showing the distributions
of the best-fit $w$, from 200 mock data sets with an inherent
$\delta_D=0.025$, after
marginalizing over the WMAP $w$CDM 5-Year priors. The shaded histogram represents the
case where $\delta_D$ is allowed to vary freely, while
for the hatched histogram the two-population systematic is ignored
($\delta_D=0$).
The dot-dashed line corresponds to the case where the data is fit assuming a
$\delta_D=0.025\pm0.025$ prior. For clarity
we omit the underlying histogram.}
\label{fig:1}
\end{figure*}

\section{Two SN Populations and the Hubble Diagram}
\label{sec:basics}

The use of SNe as standardizable candles to constrain the dark
energy EOS is based on the fundamental assumption that the lightcurves
of all individual SNe can be calibrated.  The lightcurve shape-intrinsic
luminosity relation is derived using low-$z$ samples of SNe, and it is
assumed that this relation is applicable to the higher-$z$
population.  However, if the SN population is non-uniform with
redshift, the potential for systematics must be
considered~\citep{howell:01,mannucci:06, scannapieco:05,sullivan:06}.

According to the two-population model of \citet{scannapieco:05} (henceforth SB), the prompt SNe track the instantaneous star formation rate,
$\dot{M_{*}}(t)$, and dominate the total SN rate at early
times (high redshifts) when star formation is more active.  The rate of the
delayed component scales  proportionally to the total stellar mass at a given instant, $M_{*}(t)$, 
and thus delayed SNe dominate at late times (low redshifts). The total SN rate
can be written as
\begin{equation}
\frac{SNR_{\mbox{\footnotesize Ia}}(t)}{(100\,\mbox{yr})^{-1}} = A
\left[\frac{M_{*}(t)}{10^{10}M_{\sun}}\right] + B
\left[\frac{\dot{M_{*}}(t)}{10^{10}M_{\sun} \mbox{Gyr}^{-1}}\right]
\label{eqn:twopop}
\end{equation}
where $A$ and $B$ are dimensionless constants. We take the
star formation rate proportional to $e^{-t/(2 Gyr)}$ \citep{mannucci:06}.
This model is likely to be a simple approximation
to a true, smooth underlying distribution of delay
times~\citep{totani:08,greggio:08,pritchet:08}. For simplicity we restrict
ourselves to the two-population model, although in \S\ref{s:future_data} we
discuss sensitivity to this assumption.

We allow for a redshift-independent residual difference ($\Delta{L}$) in the calibrated absolute luminosities
of the prompt (${L}_P$) and delayed (${L}_D$) SNe: $L_{P} = L_D + \Delta L$.  We take the delayed and prompt fractions of
the total SN population to be $f_D(z)$ and $f_P(z)$, respectively
($f_D(z)+f_P(z)=1$).

The average of the distance moduli of
all the SNe in a given redshift bin can be written as
\begin{equation}
\langle(m-M)\rangle = 5\log_{10} \left(\frac{d_L}{\mbox{Mpc}}\right) + 25 -
\bigg \langle 2.5  \log_{10} \left(\frac{L}{{L}_{ref}}\right)\bigg \rangle \, ,
\end{equation}
where $L_{ref}$ is the reference luminosity corresponding to
the reference absolute magnitude $M_{ref}$ given by
$M_{ref}=\bar{f}_DM_D+\bar{f}_PM_P$, where $\bar{f}_i$ is the
redshift-average of $f_i(z)$ over the low redshift range where the
calibration is done.
The residual systematic correction to $\langle (m-M)\rangle$ can be expressed as 
$\langle(m-M)\rangle_{\mbox{\scriptsize res}} = \delta_D f_D(z) - 2.5
 \log_{10} ({L}_P/{L}_{ref})$, where $\delta_D=1.086\ln(1+\Delta L/L_D)$ is the two-population bias. Note that one can also express $\langle(m-M)\rangle_{\mbox{\scriptsize res}}$
relative to the prompt component, using $\delta_P$ instead of  $\delta_D$, and $f_P$ instead of $f_D$. This involves an overall sign change, but the final results are unaltered.

Using the  $\chi^2$-statistic, we fit the Hubble diagram with
a  modified form of the distance
modulus which includes a possible residual:
\begin{equation}
(m-M)^{\rm fid}(z) = 5 \log_{10} \left(\frac{d_L(z)}{\mbox{Mpc}}\right) + 25 +
\mathcal{M} + \delta_D f_D(z).
\label{eq:dist_mod_fid}
\end{equation}
We have absorbed the redshift-independent term in 
$\langle(m-M)\rangle_{\mbox{\scriptsize res}}$ 
into the ``nuisance parameter", $\mathcal{M}$.
Note that $H_0$ is incorporated into $d_L$ instead of $\mathcal{M}$.
The redshift-dependent factor $f_D(z)$ can be
determined based on our knowledge of the star-formation
history, and $\delta_D$ must be estimated directly
from the SN data. 
This residual systematic must
now be marginalized over, and will affect cosmological parameter
estimates.

\begin{figure*}[!t] 
\centerline{
\includegraphics[scale=0.5]{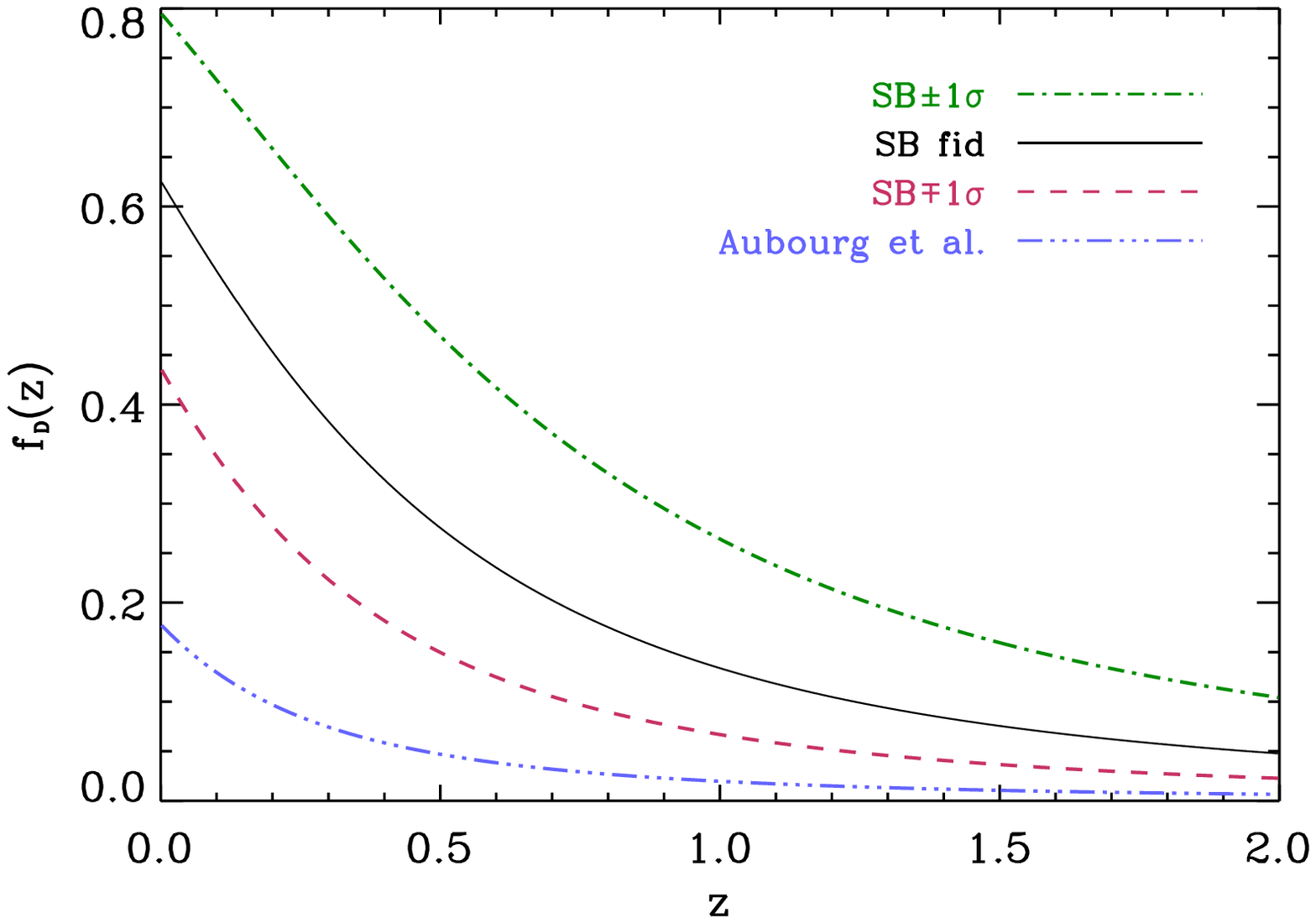} 
\includegraphics[scale=0.5]{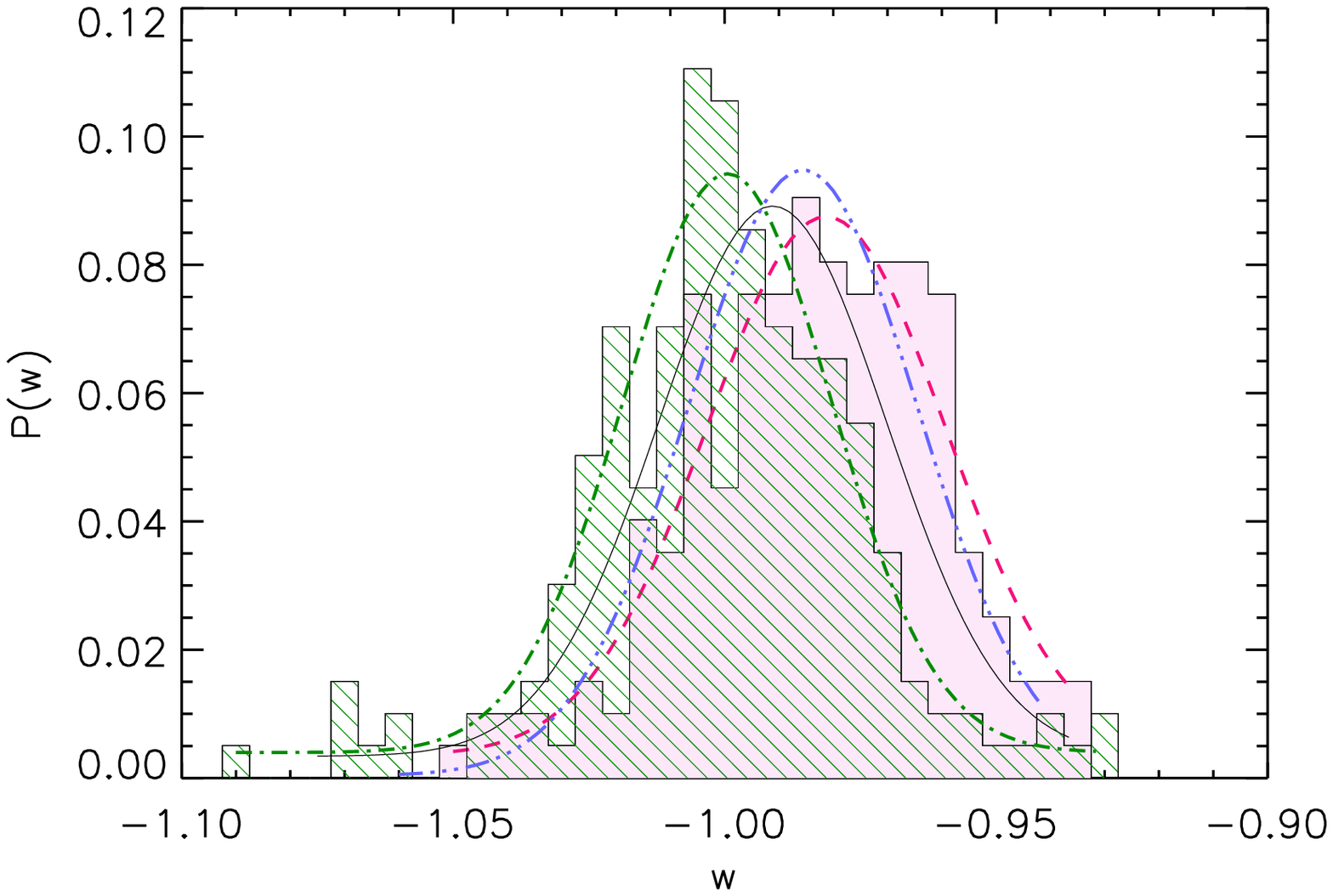}
}
\caption{{\it Left panel:} The fraction of the delayed component, $f_D$,   
as a function of redshift. The
solid line corresponds to the fiducial $A$ and $B$ values from SB~\citep{scannapieco:05} that are used to
generate the mock catalogs. The dashed and dot-dashed lines correspond to
$\{A-\Delta A, B+\Delta B\}$ and $\{A+\Delta A, B-\Delta B\}$, respectively,
where $\Delta A$ and $\Delta B$ are the $1\sigma$ uncertainties on $A$ and $B$
from SB. The triple-dot-dashed line depicts the recent estimate of $f_D$ from
\citet{aubourg:07}.
{\it Right panel:} Histograms showing the best-fit values for the dark energy   
EOS parameter, $w$, for 200 mock SN datasets with $\delta_D=0.025$. The solid curve shows
the Gaussian fit to $P(w)$ for the case where we use the same $f_D(z)$ to
generate the mock catalogs and fit the data.
The other Gaussian curves correspond to the best-fit values of $w$ when we fit
the mock data with the wrong functional form for $f_D(z)$. The mocks are
generated with the fiducial model, and the line-types represent the model used
in the fit, in accordance with the left panel. For clarity we only show two
underlying histograms, for the cases
$\{A-\Delta A, B+\Delta B\}$ (shaded) and $\{A+\Delta A,B-\Delta B\}$ (hatched).
A $\sim30\%$ shift in $f_D$ (left panel) corresponds to a percent-level bias in
$w$ (right panel). The extreme model of \citet{aubourg:07} corresponds to an
almost redshift-independent $f_D$ (see left panel), and thus the bias is similar
to that in the $\delta_D=0$ case.}
\label{fig:2}
\end{figure*}

\section{Residual systematic and parameter estimation}

\subsection{Existing data}

To study the extent to which existing data may be affected by a
residual systematic in the luminosity difference between prompt and
delayed SNe, we model fit a combined data set of 192 SNe 
\citep{davis:07, wood-vasey:07, riess:07, snls, hamuy:96a, Riess:99,
jha:06}. We also include two baryon acoustic
oscillation (BAO) distance estimates at $z=0.2$ and $z=0.35$
\citep{percival:07}, and the dimensionless distance to the surface of
last scattering $R=1.710\pm0.019$ \citep{komatsu:08}.  We take a flat
$\Lambda$CDM cosmological model and marginalize over $\mathcal{M}$,
with {\it WMAP} priors of $H_0=71.9\pm2.6$ and $\Omega_m
h^2=0.1326\pm0.0063$~\citep{komatsu:08}. These priors are independent of the SN data.
For simplicity we do not incorporate the correlation between
$\Omega_m h^2$ and $H_0$. Such a correlation will not qualitatively alter our
results, but will need to be taken into account once precision data becomes available. 

The distribution function for $\delta_D$ from existing data has a positive mean
value ($\delta_D=0.049$), implying that the 
post-calibration luminosity of the delayed population is dimmer than
the prompt population by $\sim$5\%. The standard deviation of this $\delta_D$ is
$\sigma=0.097$, which suggests that current SN data is consistent with the absence of a
two-population bias. The \citet{howell:07} result of $\sim$ $(12 \pm 4)$\% difference in the intrinsic luminosity is a {\em
pre}-calibration difference.  We find a $\sim(5 \pm 9)$\%
difference in the {\em post}-calibration luminosity between the two
types of SNe.  While this uncertainty
is larger than the pre-calibration value, it is estimated
directly from the Hubble diagram, and is independent of the empirical stretch-luminosity relation. This
important consistency check will improve as the SN sample sizes
increase.

To study the impact of a potential luminosity difference on the
measurement of the dark energy EOS, we also fit the same data (SN+BAO+CMB) 
assuming a $w$CDM model, and take the corresponding {\it
WMAP+HST}\/ priors $H_0=72.1\pm7.5$ and $\Omega_m
h^2=0.1329\pm0.0066$~\citep{komatsu:08,freedman:01}. We first consider two extreme
cases: ignoring the two-population systematic ($\delta_D=0$), and
allowing for a completely unconstrained systematic ($\delta_D$ with no
priors). With $\delta_D$ free, our analysis yields a
best-fit time-independent EOS parameter $w=-0.986 \pm 0.180$, with a
best-fit $\delta_D = 0.040 \pm 0.279$.  
If we set $\delta_D=0$, the same data provides a constraint of $w=-0.960 \pm
0.066$. By incorporating a two-population
effect in the fit, the errors on the best-fit EOS degrade by a
factor of $\sim3$. In addition, we note a shift in
the best-fit value of $w$ between the two cases.  {\em The evolution
in the ratio between prompt and delayed SNe, as a function of
redshift, can mimic dark energy.}  If there is a residual difference
in the calibrated luminosity, ignoring $\delta_D$ might bias the dark
energy estimate.  As we show in the left panel of Figure~1,
the increase in the error of the best-fit EOS is due to the degeneracy
between dark energy ($w$) and the two-population systematic ($\delta_D$).

Following \citet{howell:07}, it may be possible to
calibrate out the luminosity difference between the two components
with large samples of SNe (e.g., by looking for characteristic
properties in the SN spectra, or by correlating SN luminosities with
host galaxy types). This will lead to a prior constraint on
$\delta_D$, although uncertainties will still remain on the redshift
evolution (as estimated by $f_D(z)$). Assuming the two-population
fraction, and its redshift evolution, is perfectly known, we analyze
the same SN data assuming a Gaussian prior on $\delta_D$. 
Taking this prior to have zero mean and $\sigma=
(0.25, 0.1, 0.05)$, the errors on $w$ are ($0.130, 0.086, 0.072$), 
with the best-fit value of $w$ being approximately
the same as was found for the $\delta_D=0$ case.  Thus, by including a
two-population systematic in the fit, with a prior on $\delta_D$
centered at 0 with a 0.05 mag dispersion, we recover the same best-fit
$w$, but with $\sim10\%$ degradation in the error bars.

\subsection{Future data}
\label{s:future_data}

We now turn to proposed SN surveys, and investigate how uncertainties
in a possible two-population systematic impact measurements of $w$.  We generate mock SN catalogs with 300 SNe
uniformly distributed at $z<0.1$, and 2000 SNe in the range $0.1 \leq
z \leq 1.7$, similar to a {\it JDEM}\/-like survey \citep{kim:04}.  We
incorporate an intrinsic Gaussian scatter of 0.1 mag for each SN, and
take the relative fraction of delayed and prompt SNe, $f_D(z)$, given
by Eq.~\ref{eqn:twopop} with SB values of $A=4.4\times 10^{-2}$ and
$B=2.6$.  We assume different values (0.025, 0.05, 0.1 mag) for the underlying two-population bias ($\delta_D$).

We fit each mock data set of 2300 SNe, along with the 2 existing BAO
measurements, to a $w$CDM model, with the corresponding $w$CDM {\it WMAP+HST}\/ priors as before.
When fitting the data we consider two extreme cases: $\delta_D=0$, and $\delta_D$
completely unconstrained. Our results for $\delta_D=0.025$, from
200 separate mocks, are summarized in the right panel of Figure~1.  The
hatched histogram, which peaks at -0.974, depicts the case where
$\delta_D=0$ was assumed in the fit. This corresponds to a systematic
being present in the data, but ignored in the fit.  With an average
error on the EOS of 0.029 from MCMC, the resulting bias in the
best-fit EOS value is $\sim1\sigma$ from the underlying value ($w=-1$).

The shaded histogram in the right panel of Figure~1, having a peak at
-0.991 with a $1\sigma$ width of 0.071, shows the distribution of the
best-fit $w$ with 200 mocks, where we let $\delta_D$
vary freely while fitting the data to Eq.~\ref{eq:dist_mod_fid}.  In
this case we find no significant bias ($w=-1$ is recovered within
$<0.1\sigma$), but the width of the distribution and MCMC results for
each mock sample show that the errors in the best-fit $w$ increase by a
factor of $\sim2.5$. When the underlying two-population bias is
$\delta_D=0.05$ mag, we find $w=-0.955 \pm 0.030$ (assuming
$\delta_D=0$ in the fit) and $w=-0.986 \pm0.067$ (assuming $\delta_D$
unconstrained) based on 50 Monte-Carlo realizations. Generating data
with $\delta_D=0.1$, and neglecting the systematic in the fit, we find $w=-0.925 \pm 0.029$ (50
realizations). As a rough rule, if the two-population systematic is
neglected, the resulting bias in the best-fit $w$ is on the order of
the magnitude of the underlying $\delta_D$.

Thus far we have assumed no prior knowledge on the values of
$\delta_D$, although it may be possible to adduce {\it a priori} constraints on
$\delta_D$ through SN population statistics combined with correlations to
galaxy properties. We model-fit 200 separate SN mocks 
 (with intrinsic $\delta_D=0.025$ mag) with two
different priors: $\delta_D = 0.025\pm0.025$ and $\delta_D =0\pm0.025$ mag. The
former case represents knowledge of the true underlying systematic (with a
$1\sigma$ uncertainty of 0.025 mag), while for the latter case the
central value is incorrectly assumed to be zero (with a $0.025$ mag
uncertainty). With the $\delta_D$ prior peaked on the correct
value (0.025), the distribution of $w$ peaks at $-0.993$
with a $1\sigma$ width of 0.034 (the Gaussian with dot-dashed line in the right panel of Figure~1). 
With the prior centered on the wrong value (0 instead of 0.025), the distribution peaks
at $w=-0.977$ (showing a small bias of $\sim0.6\sigma$), with the same
$1\sigma$ uncertainty of 0.034.  In both cases the errors in $w$
increase by $\sim30\%$, when compared to the equivalent dataset with
no two-population effect in either the mock data or the fit.

Thus far we have assumed that we know $f_D(z)$ perfectly. Even if the two-population
model is correct, uncertainties in $A$ and $B$, or equivalently uncertainties in
the star-formation history,
lead to a redshift-dependent uncertainty in $f_D(z)$. To test the effect of these uncertainties on
parameter estimation, we generate mock data sets
with $\delta_D=0.025$, and $f_D(z)$ taken to be the canonical two-parameter
form given in Eq.~\ref{eqn:twopop}. We then fit these data assuming different
forms for $f_D(z)$ obtained by accounting for uncertainties in $A$ and $B$ from
SB. We also consider an estimate of $f_D$ from \citet{aubourg:07}. The
resulting $f_D$ curves are shown in the left panel of Figure~2. In the right
panel we show the resulting bias in dark energy EOS measurements, as a result of
the uncertainty in $f_D$. The uncertainties in the population fraction lead to biases in the
resulting EOS parameters. To control this bias to the percent level, the underlying distribution must be
characterized to $\lesssim20\%$.

In summary, we have found that a post-calibration shift
in the standard-candle brightness between delayed and prompt SNe can
introduce bias in the best-fit dark energy parameters. By controlling
the magnitude of any resulting two-population difference to better
than 0.025 mag, the bias can be kept under 1$\sigma$ for a {\it
JDEM}\/-like survey without significantly degrading the accuracy of
the dark energy measurements.

\acknowledgements We thank Henry Ferguson, Saurabh Jha, Mario Livio,
Adam Riess, Mark Sullivan, and Michael Wood-Vasey for useful
discussions. We thank the referee, Alex Conley, for his extremely prompt and useful comments.
This work was supported  by IGPP Astro-1603-07 at UCI and LANL, NSF CAREER 
AST-0645427, and a McCue Fellowship (AA). AC and DEH thank the Aspen
Center for Physics for hospitality while this research was completed.

%\clearpage

\end{document}